\documentclass{article}%
\usepackage{amsmath}
\usepackage{amsfonts}
\usepackage{amssymb}
\usepackage{graphicx}
\usepackage{float}
\usepackage[font=small,labelfont=bf]{caption}%
\providecommand{\U}[1]{\protect\rule{.1in}{.1in}}
\begin{document}

\begin{center}
\bigskip Lorentz Violation: Loop-Induced Effects in QED and Observational Constraints

\bigskip

\bigskip

{\Huge \bigskip}

\textbf{Z. Kepuladze}\footnote{zurab.kepuladze.1@iliauni.edu.ge,
zkepuladze@yahoo.com}

\bigskip

\textit{Institute of Theoretical Physics, Ilia State University, 0162 Tbilisi,
Georgia\ \vspace{0pt}\\[0pt]}

\textit{and} \textit{Andronikashvili} \textit{Institute of Physics, 0177
Tbilisi, Georgia\ }

\bigskip

\bigskip

\textbf{Abstract}
\end{center}

Lorentz invariance is a cornerstone of modern physics, yet its possible
violation remains both theoretically intriguing and experimentally
significant. In this work, using quantum electrodynamics as an example, we
explore how Lorentz invariance violation, introduced into a specific sector of
the theory, spreads through loop corrections, modifying the propagation and
dispersion relations of other particles. The self-energy and vacuum
polarization graphs reveal how LIV effects transfer across sectors,
influencing particle kinematics. Due to these loop effects, constraints from
cosmic-ray observations and other Earth-based experiments impose limits on
induced LIV parameters that would otherwise be less constrained. We show that
while interaction-based LIV effects require unrealistically large parameters
for detection, modifications to dispersion relations can be probed down to
$\delta\sim10^{-8}-10^{-9}$ at the LHC. This suggests that accelerator-based
resonance studies provide a promising avenue for stringent LIV constraints,
potentially rivaling astrophysical observations.

%%%%%%%%%%%%%%%%%%%%%%%%%%%%%%%%%%%%%%%%%%%%%%%%%%%%%%
\thispagestyle{empty}\newpage

\section{\bigskip Introduction}

Lorentz invariance (LI) is a foundational symmetry of the Standard Model and
of high-energy physics in general. Because it is so deeply embedded in our
theoretical framework, any deviation from exact LI --- Lorentz invariance
violation (LIV) --- would have profound implications for our understanding of
nature. Testing this symmetry's limits is therefore a central goal, both as a
precision check of the Standard Model and as a possible window into new physics.

While many LIV searches have placed extremely tight bounds --- for example, in
the photon, proton and electron sectors --- other sectors remain much less
constrained. High-energy astrophysical phenomena have long been discussed in
this context. The AGASA observations of ultra-high-energy cosmic rays beyond
the Greisen--Zatsepin--Kuzmin cutoff \cite{AGASA, GZK, Auger} and the timing
of neutrinos from \emph{Supernova 1987A} have both been interpreted, in some
analyses, as potentially allowing for LIV effects in propagation
\cite{Neutrino}. Threshold reactions such as photon decay \cite{phdecay},
vacuum Cherenkov radiation \cite{cher rad}, and synchrotron radiation
\cite{Sync}, provide remarkably precise constraints in some channels, yet
leave significant room for measurable LIV in others. If we are to take a look,
the general tendency is that LIV modifications in dispersion relations are
much more tightly constrained than those in interactions
\cite{Kostelecky-Russell}. These are often expressed in terms of a shift in
the maximal attainable velocity $c$ for a given particle species --- sometimes
colloquially referred to as the \textquotedblleft speed of
light\textquotedblright\ for that species. Following the Coleman--Glashow
parametrization \ \cite{Colleman-Gleshow}, such a modification can be written
as $E^{2}-(1-\Delta c/c)P^{2}=m^{2}$, with $\Delta c/c$ quantifying the
deviation. In this work, we will compare our calculations with the existing
experimental limits on $\Delta c/c$.

From a broad viewpoint, there are two qualitatively distinct ways LIV could
appear at high energy. Either (i) Lorentz invariance is an emergent low-energy
approximation that fails at high energy \cite{Nielsen-Chadha}, analogous to
the relation between relativistic and non-relativistic mechanics (Newtonian
mechanics); or (ii) Lorentz invariance is exact in the ultraviolet but becomes
broken at lower energies (for example via spontaneous breaking) \cite{gauge1,
Chkareuli-kep0, Chkareuli-Kep}, in which case the violation may be largely
superficial and manifest only as small effects, leaving the low-energy theory
effectively Lorentz invariant. These broad possibilities motivate the
effective field theory (EFT) approach adopted here: introduce representative
LIV operators, analyze their consequences, and compare them with experimental
constraints, without committing to a specific ultraviolet completion.

A crucial point is that LIV need not be uniform among sectors. Even if
initially identical at some high scale, LIV parameters will generally evolve
differently under renormalization group running. In gauge theories, this
running is closely tied to gauge invariance: if exact GI is absent, LIV
effects can be generated or enhanced in specific
operators\footnote{\emph{While the cited works \cite{run} do not directly
address the interplay between LIV and gauge invariance, they do examine
general features of renormalization and beta-function running in different,
simplified scenarios. These studies highlight that the subject is vast yet
incomplete, and thus motivate our choice to emphasize general logical
conclusions rather than attempt an exhaustive treatment here.}
\par
\emph{{}}}. Conversely, a nominally GI LIV operator can, through loop effects,
induce gauge-violating structures. This special relation between GI and LIV
has been discussed in \cite{gauge1, Chkareuli-kep0, gauge2} and will be
examined in more detail in Sec. 2. As a result, one may expect a hierarchy
among LIV operators at low energies, with a single operator often dominating
observable effects. This motivates our strategy of analyzing different
operators independently in distinct \textquotedblleft
what-if\textquotedblright\ scenarios.

The idea that LIV effects can be transferred between particles and sectors
through loop corrections has also been explored in earlier works. For example,
in \cite{Jackiw-Kostelecky}, a Chern--Simons term is induced from the vacuum
polarization diagram via a specific mass-type LIV operator for the fermion. In
\cite{Loop, Satunin}, various loop effects are analyzed for LIV operators
introduced at tree level in GI form. However, as noted above and discussed in
more detail in the following section, in the presence of LIV exact GI cannot
generally be maintained --- the two are intertwined, and one typically implies
the breaking of the other. In practical terms, when GI is absent and a single
LIV coefficient governs both interaction vertices and kinematic modifications,
the resulting observables combine these effects inseparably: one cannot
unambiguously attribute the deviation to either the vertex or the propagator
alone. Moreover, in many of these works, the photon polarization loop integral
is evaluated primarily in the regime $\frac{m_{fermion}^{2}}{k_{\mu\text{
}ingoing}^{2}}>1$, leaving the high-energy scaling behavior unexplored.

Notably, violations in the kinematic sector can alter the dispersion relations
of intermediate particles, opening novel possibilities for detecting LIV in
resonance energy regions at high-energy colliders. At the current
highest-energy collider, the LHC, many scattering processes proceed via
massive intermediate bosons. Since Higgs physics remains relatively new and
many related measurements lack precision, studying the weak boson resonance
region for possible LIV signatures may be a promising approach in this context.

So, in sec. 2 we introduce the operators and compute loops. In sec. 3 we
discuss collider implications and finally we conclude in sec. 4.

\section{Calculation of the loop correction}

In this chapter, we aim to calculate the "secondary effects" of several LIV
operators using quantum electrodynamics (QED) as an example. Let us discuss
massless electrodynamics in the Feynman gauge. While massless QED is the
technically simpler framework, the qualitative conclusions extend to a wider
class of gauge theories. For instance, the fine structure constants of QED and
weak interactions are very close to each other. In loop calculations at high
energies, mass plays only a secondary role. Clearly, the massless case may
spoil the infrared limit; however, when studying LIV effects, the infrared
limit is not our primary concern. At the same time, we still provide insight
into previously unexplored high-energy behavior. Before we proceed with this
plan, let us expand on the relation between LIV and gauge invariance in a bit
more detail.

\subsection{LIV vs GI}

When considering LIV modifications in a GI setup, it is also important to
specify the gauge. If the gauge is not fixed, any LIV operator can effectively
serve as a gauge-fixing term and leave the theory Lorentz invariant. A simple
example is the mass term $m^{2}(n_{\mu}A^{\mu})^{2}$ inserted into an
otherwise GI Lagrangian.

Here $n_{\mu}$ denotes a fixed background unit vector that selects a preferred
spacetime direction. We normalize it as $n_{\mu}n^{\mu}=\pm1$, depending on
whether the vector is time-like ($n_{\mu}=(1,0,0,0)$) or space-like ($n_{\mu
}=(0,0,0,1)$). In what follows, we treat $n_{\mu}$ as a constant vacuum
background in the sense that it does not vary from point to point in
spacetime. However, it transforms as a Lorentz vector and only takes the
simple component form in a preferred frame, typically associated with the
reference frame of distant stars. By contrast, treating $n_{\mu}$ as
non-transforming would lead to an unphysical picture.

In the example above, one could perform a gauge transformation to the axial
gauge $A_{\mu}n^{\mu}=0$ to eliminate the mass term. In this case, the LIV
operator simply fixes the gauge, and the resulting theory remains perfectly
Lorentz invariant. The same reasoning applies to interaction terms such as
$(A_{\mu}n^{\mu})\overline{\Psi}{\not n  }\Psi$ and $(A_{\mu}n^{\mu}%
)\overline{\Psi}\Psi$, where by convention ${\not n  }=n_{\mu}{\gamma}^{\mu}$.

More generally, if one introduces operators $F(A_{\mu},n_{\nu})$, $F(A_{\mu
},n_{\nu})\overline{\Psi}{\not n  }\Psi$ or $F(A_{\mu},n_{\nu})\overline{\Psi
}\Psi$, and the corresponding gauge condition $F(A_{\mu}+\partial_{\mu}%
\omega,n_{\nu})=0$ admits a solution for the transformation parameter $\omega$
for arbitrary $A_{\mu}$, then the operator amounts to gauge fixing and does
not generate genuine LIV. If no such solution exists, the theory lacks exact
gauge invariance and the LIV effects are governed by the non-gauge invariant operator.

A distinct case arises when $F(A_{\mu},n_{\nu})$ itself has a GI form, e.g.
$F(A_{\mu},n_{\nu})\sim n_{\mu}F^{\mu\lambda}n^{\nu}F_{\nu\lambda}$, with
$F^{\mu\lambda}=\partial^{\mu}A^{\lambda}-\partial^{\lambda}A^{\mu}$. This
corresponds to an extension of the Lagrangian,
\begin{equation}
\Delta L=\delta n_{\mu}F^{\mu\lambda}n^{\nu}F_{\nu\lambda} \label{T1}%
\end{equation}
with $\delta$ a LIV parameter, presumably small, but not necessarily so.

For a massless vector field, this modification does not spoil the
Lorentz-invariant limit: the field still propagates two degrees of freedom on
shell and consistently reproduces the Coulomb law. No apparent contradiction
arises at this stage. The situation changes once U(1) symmetry is
spontaneously broken in the presence of (\ref{T1}). The vector field then
acquires a mass term. Because the setup retains a formally GI structure, one
can go to the unitary gauge and calculate the propagator of the massive vector
field. There, it becomes clear that the massive vector is over-constrained:
instead of three propagating degrees of freedom, as in the LI case, the field
still carries only two. This fails to describe a realistic massive vector particle.

The same conclusion follows if we consider two abelian gauge fields with the
operator (\ref{T1}) and break symmetry using a complex scalar. One linear
combination becomes massive, while the orthogonal combination remains
massless, yet both the massive and massless fields still propagate only two
degrees of freedom each. Thus, the degree-of-freedom mismatch persists. The
massive vector field still appears over-constrained. We can straightforwardly
expand this logic to the electroweak symmetry of the Standard Model, where
massive modes again lack the necessary degrees of freedom, and mixing of
different fields does not cure the deficit. In summary, while in the unbroken
phase the operator (\ref{T1}) shows no immediately obvious inconsistency, once
the symmetry is broken the problem becomes explicit: the vector acquires a
mass term yet continues to describe only two propagating degrees of freedom
instead of three, inconsistent with the structure of a realistic massive
vector boson.

This occurs because LIV and GI are so deeply intertwined. In fact, LIV is
obscured by gauge degrees of freedom. In \cite{gauge1, Chkareuli-kep0}, it is
demonstrated how physical LIV is directly correlated with the size of GI
violation. Further research in this matter indicates that gauge theories, both
Abelian and non-Abelian, can be obtained by themselves from the requirement of
the physical non-observability of the spontaneous LIV \cite{gauge2}. But one
may ask, how about LIV operators not directly connected to the vector fields?
For example LIV mass type of terms, similar to one considered in
\cite{Jackiw-Kostelecky}, $\overline{\Psi}\not b  \,\gamma_{5}\Psi$ with some
prescribed constant vector $b^{\nu}$. This term is invariant under the global
U(1) transformation and henceforth appears manifestly gauge invariant on the
surface. However, it breaks not only Lorentz invariance, but also GI as well
and this becomes apparent when calculating photon polarization loop taking
this modification into account. If $\Pi^{\mu\nu}(p;b)$ denotes the modified
photon polarization with external momentum $p_{\lambda}$, then expanding in
powers of $b^{\nu}$ gives
\[
\Pi^{\mu\nu}(p;b)\approx\Pi_{0}^{\mu\nu}+\Pi_{b}^{\mu\nu}+\Pi_{bb}^{\mu\nu}%
\]
$\ $where $\Pi_{0}^{\mu\nu}$ is the standard GI ($\Pi_{0}^{\mu\nu}p_{\mu}%
=0$)\ result. $\Pi_{b}^{\mu\nu}$ corresponds to a Chern--Simons--like term,
which can appear GI in form ($\Pi_{b}^{\mu\nu}p_{\mu}=0$), but is known to be
regularization dependent \cite{Jackiw-Kostelecky}, while $\Pi_{bb}^{\mu\nu}$
is no longer gauge invariant, $p_{\nu}\Pi_{bb}^{\mu\nu}\neq0$ (See appendix A
for proof).

However, one may ask: what does it matter if GI is only broken at second order
in $b^{\nu}$, since such effects might seem practically irrelevant? The
important point is that already at linear order the situation is problematic:
the induced Chern--Simons term has no unique, regulator-independent value.
This ambiguity is well documented in the cited work and subsequent literature
\cite{Ind}. We focused on the self-energy loop diagram because it is central
to many discussions of LIV in QED.

Case of the mass term $\overline{\Psi}\not b  \,\Psi$ is even simpler. In an
otherwise GI setup this term is gauged away by selecting the transformation
parameter to be $b_{\mu}x^{\mu}$, once again proving the point. One could try
to explicitly check higher-order tensor and/or pseudo-tensor structures for
fermions, but this would take us farther away from the main goal of the paper,
especially when the provided examples paint the picture vividly.

One must state that the same LIV operator may have different effects in
differently fixed gauges; they might not be equivalent, and there might not
exist any LIV observable that remains unchanged when moving from one gauge to
another. This should not be a surprise, since for LIV to occur, GI violation
is necessary, and therefore frameworks with different gauge conditions are not
physically equivalent\footnote{\emph{One might draw a parallel with finite-temperature
QFT, which also introduces a preferred rest frame and therefore breaks LI
while remaining exactly gauge invariant. However, the analogy has important
limitations. Thermal LIV effects originate from ensemble averages of physical
degrees of freedom and arise from loop corrections performed in a fixed gauge.
These effects are physical and gauge-invariant: the solution (thermal
background) lacks the symmetry of the equations, but the gauge structure
remains intact. In contrast, in spontaneous LIV the fixed vacuum expectation
value is a non-invariant classical solution that does not distinguish between
gauge and physical modes. When physical degrees of freedom are isolated, the
gauge modes can absorb the would-be LIV, rendering it unobservable unless GI
is also broken. Thus, thermal LIV and spontaneous LIV have similar appearances
but fundamentally different mechanisms.}
\par{}}. Perhaps fixing the gauge in LIV scenarios requires some
hard motivation itself, similar to the axial gauge motivated by spontaneous
LIV. Whether the Feynman gauge is the optimal choice or is physically the most
motivated option is an entirely separate discussion. However, let us abandon
that discussion as it is beyond the scope of this paper.

Adopting LIV represents a significant conceptual leap, but quantitatively
these effects are expected to be extremely small and exotic. Therefore, any
phenomenological model must account for every subtlety, however inconvenient.
This is equally crucial experimentally, since data analysis relies heavily on
model fitting and any potential LIV signal demands exceptionally careful
interpretation given its anticipated small magnitude.

In the framework of massless quantum electrodynamics in the Feynman gauge, let
us first analyze the case of LIV interaction and investigate its influence
through loop effects on the photon and fermion propagators. Subsequently, we
modify the dispersion relation of the photon and the electron and examine
their mutual influence on each other. We do not attempt to calculate LIV
corrections to the effective vertex, since this part of electrodynamics is the
least constrained and cannot provide practically useful limits on LIV parameters.

\subsection{LIV interaction}

In this section, we first consider the specific scenario in which the
leading-order LIV operator modifies the interaction term. Many types of
photon--fermion interactions that violate Lorentz invariance could be
introduced once higher-dimensional operators are allowed. However, if the LIV
background is fixed by a unit vector $n_{\mu}$ and we restrict ourselves to
dimension-four operators, only three independent possibilities remain. Two of
these were already mentioned above, $(A_{\mu}n^{\mu})\overline{\Psi}{\not n
}\Psi$ and $(A_{\mu}n^{\mu})\overline{\Psi}\Psi$, while the third is $(A_{\mu
}n_{\nu})\overline{\Psi}\sigma^{\mu\nu}\Psi$ \footnote{\emph{Strictly
speaking, this classification is not complete. The pseudo-tensor counterparts
of these operators are also admissible. However, in massless QED they do not
generate linear contributions in two-point loop integrals, and are therefore
irrelevant for the present analysis.}}. The operator\ $(A_{\mu}n^{\mu
})\overline{\Psi}\Psi$, while the simplest, is not very characteristic of
vector--fermion interactions. The operator $(A_{\mu}n_{\nu})\overline{\Psi
}\sigma^{\mu\nu}\Psi$ resembles a structure typically induced by loop effects,
similar to the anomalous magnetic moment, and is therefore expected to be
suppressed compared to a primary LIV source. In contrast, the operator
$(A_{\mu}n^{\mu})\overline{\Psi}{\not n  }\Psi$ has a form consistent with
spontaneous LIV. If $n_{\mu}$ represents the vacuum expectation value
associated with spontaneous LIV, then $A_{\mu}n^{\mu}$ can be interpreted as
the Higgs-like mode interacting with the fermion current projected onto the
vacuum direction. Thus, it is not an exaggeration to claim that the simplest
and physically most motivated choice is%
\begin{equation}
\Delta L_{int}=e\delta_{\text{int}}(A_{\mu}n^{\mu})\overline{\Psi}{\not n
}\Psi\label{int1}%
\end{equation}

The restriction on $\delta_{\text{int}}$ comes from the covariant derivative
form always introduced for fermions, $c^{\mu\nu}\overline{\Psi}(\gamma_{\nu
}D_{\mu})\Psi$ to uphold GI. In this construction, $c^{\mu\nu}$ corresponds to
our $\delta_{\text{int}}n^{\mu}n^{\nu}$. For time-like violation, limits for
the electron from the clock redshift comparisons and the tests of Einstein
equivalence principle lie below $10^{-9}$ \cite{Time}, while laboratory bounds
from synchrotron and Cherenkov radiation give $4\cdot10^{-15}$ \cite{Time1}.
For space-like violation the limits are at the level of $10^{-15}$
\cite{Space}. Some more recent estimates from astrophysical sources are even
more stringent, reaching $10^{-19}$ \cite{Spacetime}. While these
astrophysical bounds are strict, they often involve assumptions about source
modeling or the photon sector, so some caution is warranted. Importantly,
these bounds arise from constraints on the electron's velocity shift $\Delta
c/c$ and in many cases assume a Lorentz-invariant photon. The limits are then
carried over to the interaction constant only because of the shared GI form.
Without GI, we do not yet have direct limits on $\delta_{\text{int}}$ itself.
This opens the possibility that, by relaxing the GI assumption, one can derive
genuinely new constraints on $\delta_{\text{int}}$\ rather than merely
inheriting those tied to $\Delta c/c$. We will explore this possibility below.

\subsubsection{Transfer of LIV to fermion kinematics}

LIV modifications arise from quantum loop effects, which can alter particle
propagation. In particular, self-energy corrections provide insight into how
LIV influences the fermion dispersion relation. We now compute the self-energy
loop to quantify these modifications. For an incoming momentum $p_{\mu}$, the
self-energy loop calculation is:
\begin{align}
\Pi &  =e^{2}%
%TCIMACRO{\tint }%
%BeginExpansion
{\textstyle\int}
%EndExpansion
\frac{d^{4}q}{(2\pi)^{4}}\left(  \gamma^{\mu}+\delta_{\text{int}}n^{\mu
}{\not n  }\right)  \frac{1}{\not q  }\left(  \gamma^{\mu}+\delta_{\text{int}%
}n^{\mu}{\not n  }\right)  \frac{1}{(p-q)_{\rho}^{2}}\label{self energy}\\
&  =-2e^{2}(\gamma^{\mu}(1+\delta_{\text{int}}n_{\rho}^{2})-2\delta
_{\text{int}}n^{\mu}{\not n  })%
%TCIMACRO{\tint }%
%BeginExpansion
{\textstyle\int}
%EndExpansion
\frac{d^{4}q}{(2\pi)^{4}}\frac{q_{\mu}}{q_{\lambda}^{2}(p-q)_{\rho}^{2}%
}\nonumber
\end{align}
Here, the photon propagator is taken in the Feynman gauge: \ $\dfrac
{-ig_{\mu\nu}}{p_{\alpha}^{2}}$. Since this integral can only be proportional
to $p_{\mu}$, we define:%
\begin{equation}%
%TCIMACRO{\tint }%
%BeginExpansion
{\textstyle\int}
%EndExpansion
\frac{d^{4}q}{(2\pi)^{4}}\frac{q_{\mu}}{q_{\lambda}^{2}(p-q)_{\rho}^{2}%
}=p_{\mu}I_{0}(p^{2})
\end{equation}
where $I_{0}(p^{2})$ is a divergent integral. Using a simple momentum cutoff
with a scale $\Lambda$, it can be computed as:
\begin{equation}
I_{0}(p^{2})=-\frac{i}{32\pi^{2}}\ln\frac{p_{\nu}^{2}}{\Lambda^{2}}%
\end{equation}
In principle, one could find an additional constant alongside the logarithm,
depending on the regularization scheme. However, such constants would only
redefine the\ $\Lambda$ scale without affecting the underlying physics. The
divergent part is typically isolated by introducing a scale $\mu$, which does
not alter the physical interpretation but reorganizes the expression:
\begin{equation}
I_{0}(p^{2})=-\frac{i}{32\pi^{2}}\left[  \ln\frac{\mu^{2}}{\Lambda^{2}}%
+\ln\frac{p_{\nu}^{2}}{\mu^{2}}\right]  =-\frac{i}{32\pi^{2}}\left[  \ln
\frac{\mu^{2}}{\Lambda^{2}}+I_{f}\right]
\end{equation}
where $I_{f}$ is the finite part. Notably, the division between divergent and
finite parts is not uniquely defined at this stage. Thus, $\Pi$ becomes:
\begin{equation}
\Pi=\frac{i\alpha}{4\pi}({\not p  }(1+\delta_{\text{int}}n_{\rho}^{2}%
)-2\delta_{\text{int}}(pn){\not n  })\left[  \ln\frac{\mu^{2}}{\Lambda^{2}%
}+I_{f}\right]
\end{equation}
where $\alpha$ is the well-known fine structure constant. Divergent
contributions can be eliminated using counter terms, leading to the modified
fermion propagator:%

\begin{align}
D_{f}  &  =\frac{i}{{\not p  }}(1+\Pi\frac{i}{{\not p  }}+(\Pi\frac
{i}{{\not p  }})^{2}+...)=\frac{i}{{\not p  }}\frac{1}{1-\Pi\dfrac{i}{{\not p
}}}\nonumber\\
&  =\frac{i\left(  1+\dfrac{\alpha I_{f}}{4\pi}\right)  ^{-1}}{{\not p
}(1+\overline{\delta}_{\text{int}}(n_{\alpha}^{2}-\dfrac{4(pn)^{2}}{p_{\alpha
}^{2}}))+2\overline{\delta}_{\text{int}}(pn){\not n  }}\text{\ \ }
\label{prof1}%
\end{align}
Here, $\left(  1+\dfrac{\alpha}{4\pi}I_{f}\right)  ^{-1}$ is the standard
renormalization factor, and $\overline{\delta}_{\text{int}}$ is
\begin{align}
\overline{\delta}_{\text{int}}\left(  \mu\right)   &  =\dfrac{\alpha I_{f}%
}{4\pi}\delta_{\text{int}}\label{delta}\\
\overline{\delta}_{\text{int}}(\mu_{1})-\overline{\delta}_{\text{int}}(\mu
_{2})  &  =\dfrac{\alpha}{4\pi}\delta_{\text{int}}\ln\left(  \mu_{2}^{2}%
/\mu_{1}^{2}\right) \nonumber
\end{align}
From here on, a bar indicates the loop-induced coefficient for any $\delta$
introduced below, as in (\ref{delta}).

From this running we observe that, at lower scales $\mu$,\ $\overline{\delta
}_{\text{int}}\left(  \mu\right)  $ is reduced in magnitude compared to its
value at the high scale $M_{LIV}$. If the highest scale is the LIV scale
$M_{LIV}$ (possibly near a GUT scale), then the one-loop contribution at lower
energies gives an asymptotic behavior: \
\begin{equation}
\overline{\delta}_{\text{int}}\left(  \mu\right)  =\dfrac{\alpha}{4\pi}%
\delta_{\text{int}}\ln\frac{M_{LIV}^{2}}{\mu^{2}}%
\end{equation}

In our simplified massless-fermion treatment an infrared sensitivity appears
in equation (\ref{delta}); however, the infrared energy region $\frac
{m_{electron}^{2}}{p_{\mu}^{2}}\gg1$ is well described by the author in
\cite{Satunin}, indicating a behavioral change in $\overline{\delta
}_{\text{int}}$, with $\overline{\delta}_{\text{int}}\sim E^{2}$. While the
discussed modification pertains to the photon dispersion relation, the general
treatment of low-energy behavior still holds. Examining equation (\ref{prof1})
we see that the fermion propagator pole is shifted by the loop-induced term,
leading to a modified dispersion relation:%
\begin{equation}
p_{\alpha}^{2}-4\overline{\delta}_{\text{int}}(pn)^{2}\approx0 \label{MF}%
\end{equation}
For a given fermion, we observe that $\left\vert \Delta c/c\right\vert
\sim\left\vert \overline{\delta}_{\text{int}}\right\vert $. The absolute
values are used because the sign of $\Delta c/c$ depends on the type of
violation and on the sign of $\overline{\delta}_{\text{int}}$. Experimental
constraints on electron velocity shifts are stringent, with the strongest
limits giving $\left\vert \overline{\delta}_{\text{int}}\right\vert <10^{-19}$
\cite{Kostelecky-Russell, Spacetime}. This implies a bound on the underlying
interaction parameter of order
\[
\left\vert \delta_{\text{int}}\right\vert <10^{-16}-10^{-17}%
\]
This represents a direct and novel constraint on the interaction
strength---one that does not rely on enforcing gauge invariance. A direct
comparison with tree-level $c^{\mu\nu}$-type operators would be misleading,
since such terms depend on gauge invariance. In contrast, our framework not
only avoids assuming gauge invariance but actively discourages it. Therefore,
the constraint derived here reflects a genuinely LIV-induced effect.

This result highlights how LIV in the interaction sector can transmit to
fermion kinematics through loops, and how dispersion-relation measurements
thus place meaningful constraints on interaction-type LIV operators. In this
sense, the present bound is a novel insight: earlier limits reflected only the
imposition of gauge invariance, whereas here the restriction emerges directly
from LIV effects constrained by electron lightspeed variation.

\subsubsection{Transfer of LIV to the photon kinematics}

Similar to the fermion propagator, analogous calculations can be performed for
the photon polarization fermion loop. If only the Lorentz-violating
interaction (\ref{int1}) is employed, the first approximation gives:
\begin{equation}
\overline{\Pi}_{\mu\nu}=\Pi_{\mu\nu}+\delta_{\text{int}}(n_{\mu}n^{\lambda}%
\Pi_{\lambda\nu}+n_{\nu}n^{\lambda}\Pi_{\lambda\mu})
\end{equation}
where $\Pi_{\mu\nu}$ is the standard expression for electrodynamics
\cite{Kaku}:%
\begin{equation}
\Pi_{\mu\nu}=(p_{\alpha}^{2}g_{\mu\nu}-p_{\mu}p_{\nu})\Pi_{0}%
\end{equation}
This allows us to identify the correction to the propagator as:
\begin{equation}
\Delta D_{\mu\nu}=\frac{-i}{p_{\alpha}^{2}}\overline{\delta}_{\text{int}%
}(n_{\mu}(n_{\nu}-p_{\nu}\frac{n^{\lambda}p_{\lambda}}{p_{\alpha}^{2}}%
)+n_{\nu}(n_{\mu}-p_{\mu}\frac{n^{\lambda}p_{\lambda}}{p_{\alpha}^{2}}))
\end{equation}

We should not find strange terms proportional to momentum appearing in the
propagator. After all, in the Feynman gauge, the propagator is not manifestly
transverse, and besides, such terms do not affect the calculation due to
charge conservation. This type of correction does not modify the dispersion
relation of the photon. To understand the exact LIV-induced modification in
the kinetic operator, one must invert this propagator. An immediate
observation suggests that it modifies Coulomb's law for time-like violation
case for example; however, this modification is identical to what the
interaction (\ref{int1}) itself introduces but with a smaller magnitude:
\begin{equation}
\Delta A_{\mu}=\Delta D_{\mu\nu}J^{\nu}=\frac{-i}{p_{\alpha}^{2}}%
\overline{\delta}_{\text{int}}(2n_{\mu}-p_{\mu}\frac{n^{\lambda}p_{\lambda}%
}{p_{\alpha}^{2}})n_{\nu}J^{\nu}%
\end{equation}
since momentum dependent part does not contribute anyhow in the
electromagnetic field.

Corresponding corrections to the kinetic operator follow a similar form to
those in the propagator, as expected:
\begin{equation}
\mathcal{K}_{\mu\nu}=i\left(  p_{\alpha}^{2}g_{\mu\nu}-\overline{\delta
}_{\text{int}}\left[  n_{\mu}(p_{\alpha}^{2}n_{\nu}-p_{\nu}\left(  n^{\lambda
}p_{\lambda}\right)  )+n_{\nu}(p_{\alpha}^{2}n_{\mu}-p_{\mu}\left(
n^{\lambda}p_{\lambda}\right)  )\right]  \right)
\end{equation}

In the Feynman gauge, the momentum-dependent part can drop out, leaving an
effective correction to the kinetic term of the form%

\begin{equation}
{\Delta \mathcal L}_{k}=2\overline{\delta}_{\text{int}}\left(  n_{\nu}%
\partial_{\lambda}A^{\nu}\right)  ^{2}%
\end{equation}

As already mentioned, this modification does not affect the photon dispersion
relation or the speed of light. Therefore, constraints from threshold-energy
arguments, Michelson--Morley--type resonator tests, rotating optical cavities,
vacuum birefringence, dispersion, or time-of-flight do not apply. Instead, the
effects arise only through off-shell photons: a small orientation dependence
in static electromagnetic interactions (e.g. LIV corrections to the Coulomb
potential for time-like violation or vector potential for general $n_{\mu}$,
and the associated spectroscopic shifts), and polarization-projected
distortions in scattering amplitudes and angular distributions.

These contributions mirror those generated by the operator (\ref{int1}), but
appear as loop-induced, parametrically suppressed versions of it. In this
sense, they inherit the same structure but with reduced magnitude.
Importantly, the interaction (\ref{int1}) itself has no direct constraints;
the bounds we applied earlier arise only through limits on electron
light-speed variation from (\ref{MF}).

\subsection{Modification of the dispersion relation of the fermion}

If, instead of the LIV interaction (\ref{int1}), the primary source of LIV is
a modification of the fermion kinetic operator,
\begin{equation}
\Delta L_{f}=\delta_{f}\overline{\Psi}{\not n  }\left(  p_{\lambda}n^{\lambda
}\right)  \Psi\label{fermion0}%
\end{equation}
then the fermion's dispersion relation is modified in a manner similar
to\ (\ref{MF}). In addition, a dynamical loop effect arises for the photon
propagator. The calculation of the photon polarization operator in this case
follows directly from the methods outlined in the previous subsection.

Using the LIV fermion propagator modified by\ (\ref{fermion0}), the resulting
one-loop perturbation of the photon propagator is:%
\begin{equation}
\Delta D_{\mu\nu}=\frac{-ig_{\mu\nu}}{p_{\alpha}^{2}}\overline{\delta}%
_{f}\frac{(p_{\lambda}n^{\lambda})^{2}}{p_{\alpha}^{2}}%
\end{equation}

Such a correction induces the following photon dispersion relation:
\begin{equation}
p_{\alpha}^{2}+\overline{\delta}_{f}(n^{\lambda}p_{\lambda})^{2}=0
\label{inducedphoton}%
\end{equation}
This form is familiar and therefore easier to constrain. Threshold constraints
in QED --- from processes such as synchrotron radiation, Cherenkov radiation,
photon decay, or pair annihilation --- primarily restrict the difference
between the photon and electron LIV parameters, $\delta_{\gamma}-\delta_{e}$,
rather than their absolute values. This difference is strongly suppressed,
with current bounds $\left\vert \delta_{\gamma}-\delta_{e}\right\vert
<5\cdot10^{-21}-5\cdot10^{-22}$ \cite{Photon}. \ 

The implication is twofold: either (i) $\delta_{\gamma}$ and $\delta_{e}$ are
of the same order, in which case both may be much larger than their tiny
difference, or (ii) there exists a hierarchy between them, with the larger
parameter saturating the bound and the smaller one being effectively
unconstrained. Interpreting our loop-induced result, the natural
identification is $\delta_{e}=\delta_{f}$ and $\delta_{\gamma}=\overline
{\delta}_{f}$. In this case, we fall into the hierarchy scenario: since
$\overline{\delta}_{f}\ll\delta_{f}$ (loop suppression), we obtain
\begin{equation}
\left\vert \delta_{e}\right\vert =\left\vert \delta_{f}\right\vert
<10^{-21}-10^{-22}%
\end{equation}
and predict a tighter bound for the photon parameter,
\begin{equation}
\left\vert \delta_{\gamma}\right\vert =\left\vert \overline{\delta}%
_{f}\right\vert <10^{-23}-10^{-24}.
\end{equation}

\subsection{LIV modification of the photon}

If the primary source of LIV originates from photon kinematics, the
modification to the Lagrangian can be written as
\begin{equation}
\Delta L_{photon}=-\frac{\delta_{ph}}{2}(n^{\mu}\partial_{\mu}A_{\lambda
})(n^{\alpha}\partial_{\alpha}A^{\lambda})
\end{equation}
This leads to the modified photon dispersion relation%
\begin{equation}
p_{\alpha}^{2}+\delta_{ph}(n^{\lambda}p_{\lambda})^{2}=0 \label{photon}%
\end{equation}
which resembles (\ref{inducedphoton}), but differs conceptually: in the
present case, LIV arises directly as a primary modification of the photon
sector, whereas (\ref{inducedphoton}) followed from loop-induced corrections
starting in the fermion sector.

Inserting (\ref{photon}) into the fermion self-energy diagram (analogous to
(\ref{self energy})), we obtain:
\begin{equation}
\Pi=e^{2}%
%TCIMACRO{\tint }%
%BeginExpansion
{\textstyle\int}
%EndExpansion
\frac{d^{4}q}{(2\pi)^{4}}\gamma^{\mu}\frac{1}{\left(  {\not p  }-{\not q
}\right)  }\gamma^{\mu}\frac{1}{q_{\rho}^{2}+\delta_{ph}(n^{\lambda}%
p_{\lambda})^{2}}%
\end{equation}

Isolating the LIV effect, we find:
\begin{align}
\Pi_{LIV}  &  =-e^{2}%
%TCIMACRO{\tint }%
%BeginExpansion
{\textstyle\int}
%EndExpansion
\frac{d^{4}q}{(2\pi)^{4}}\gamma^{\mu}\frac{1}{\left(  {\not p  }-{\not q
}\right)  }\gamma^{\mu}\frac{\delta_{ph}(n^{\lambda}q_{\lambda})^{2}}{q_{\rho
}^{2}q_{\alpha}^{2}}\nonumber\\
&  =2\delta_{ph}e^{2}\gamma^{\mu}\left[  I_{0}n_{\rho}^{2}p_{\mu}%
-2I_{1}(n^{\lambda}p_{\lambda})n_{\mu}+I_{2}p_{\mu}\frac{(n^{\lambda
}p_{\lambda})^{2}}{p_{\alpha}^{2}}\right]
\end{align}
where $I_{i}$ are loop integrals evaluated using Feynman parametrization:
\begin{align}
I_{0}  &  =\frac{1}{4}\int_{0}^{1}dx\int\frac{d^{4}q}{(2\pi)^{4}}\frac
{x^{2}q_{\rho}^{2}}{\left(  q_{\rho}^{2}+x(1-x)p_{\alpha}^{2}\right)  ^{3}%
}=\frac{-i}{384\pi^{2}}\left(  \frac{1}{2}+\frac{13}{3}\ln\frac{p_{\nu}^{2}%
}{\Lambda^{2}}\right) \\
I_{1}  &  =\frac{1}{4}\int_{0}^{1}dx\int\frac{d^{4}q}{(2\pi)^{4}}%
\frac{x(1-x)q_{\rho}^{2}}{\left(  q_{\rho}^{2}+x(1-x)p_{\alpha}^{2}\right)
^{3}}=\frac{-i}{384\pi^{2}}\left(  \frac{1}{4}+\frac{5}{3}\ln\frac{p_{\nu}%
^{2}}{\Lambda^{2}}\right) \\
I_{2}  &  =p_{\alpha}^{2}\int_{0}^{1}dx\int\frac{d^{4}q}{(2\pi)^{4}}%
\frac{x^{2}(1-x)^{2}}{\left(  q_{\rho}^{2}+x(1-x)p_{\alpha}^{2}\right)  ^{3}%
}=\frac{i}{192\pi^{2}}%
\end{align}
Here $x$ is the Feynman parameter and $\Lambda$ is the cutoff scale, as
before. Among these, $I_{0}$ and $I_{1}$ are logarithmically divergent, while
$I_{2}$ is convergent. The divergent parts can be absorbed by counterterms;
after introducing a renormalization scale $\mu$, one finds
\begin{align}
\Pi_{LIV}  &  =2\delta_{ph}e^{2}\gamma^{\mu}\left[  \left(  I_{0}n_{\rho}%
^{2}+I_{2}\frac{(n^{\lambda}p_{\lambda})^{2}}{p_{\alpha}^{2}}\right)  p_{\mu
}-2I_{1}(n^{\lambda}p_{\lambda})n_{\mu}\right] \\
&  =\frac{-i\delta_{ph}\alpha}{24\pi}\left[  \mathcal{A}(p){\not p
}-\mathcal{B}(p)(n^{\lambda}p_{\lambda}){\not n  }\right]
\end{align}
where%
\begin{align}
\mathcal{A}(p)  &  =n_{\rho}^{2}\left(  \frac{1}{4}+\frac{13}{6}\ln
\frac{p_{\nu}^{2}}{\mu^{2}}\right)  -2\frac{(n^{\lambda}p_{\lambda})^{2}%
}{p_{\alpha}^{2}}\\
\mathcal{B}(p)  &  =\frac{1}{4}+\frac{5}{3}\ln\frac{p_{\nu}^{2}}{\mu^{2}}%
\end{align}

From this, we can define the modified fermion propagator:%
\begin{align}
D_{f}  &  =\frac{i}{{\not p  }-{\not p  }\Pi_{LIV}\dfrac{i}{{\not p  }}%
}=\nonumber\\
&  =\frac{i(1-\dfrac{\delta_{ph}\alpha}{24\pi}(\mathcal{A}(p)+2(p^{\alpha
}n_{\alpha})\dfrac{\mathcal{B}(p)}{p_{\mu}^{2}})^{-1}}{{\not p  }%
+\dfrac{\delta_{ph}\alpha}{24\pi}\mathcal{B}(p)(p^{\alpha}n_{\alpha}){\not n
}}\Rightarrow\frac{i}{{\not p  }+\overline{\delta}_{ph}(p^{\alpha}n_{\alpha
}){\not n  }}%
\end{align}
thus:%

\[
\overline{\delta}_{ph}=\delta_{ph}\dfrac{\alpha\mathcal{B}(p)}{24\pi}%
\]

Once again, loop corrections transfer LIV between the photon and fermion
sectors, producing modified fermion propagation and dispersion. The structure
of the result is similar to the previous subsection, but the origin is
different. Experimental constraints on $\left\vert \delta_{\gamma}-\delta
_{e}\right\vert $ again enforce a hierarchy between photon and fermion
parameters, leading to
\begin{equation}
\left\vert \delta_{\gamma}\right\vert =\left\vert \delta_{ph}\right\vert
<10^{-21}-10^{-22}\text{, \ \ \ \ \ \ \ \ \ \ \ \ \ \ \ \ }\left\vert
\delta_{e}\right\vert =\left\vert \overline{\delta}_{ph}\right\vert
<10^{-23}-10^{-24}%
\end{equation}

In this case, however, the hierarchy is inverted relative to the scenario
where the fermion LIV was primary.

\section{About accelerator specific physics}

Experimental and observational data overwhelmingly show that modifications to
kinematics and dispersion relations are measured with far greater precision
than modifications to interactions. Beyond the constraints derived from
cosmic-ray threshold energies and other astrophysical observations,
high-energy scattering processes in accelerators provide a unique and
sensitive avenue for detecting LIV effects --- particularly when intermediate
bosons are involved. If the center-of-mass energy of an experiment is
sufficiently high, LIV-induced modifications to intermediate boson dispersion
relations become especially relevant.\ 

In the works \cite{CMS, Lunghi} LIV modifications in the quark sector were
analyzed in the context of the LHC. The modifications to cross sections depend
on the relative orientation of the scattering process (correspondingly, the
orientation of the proton--proton collision axis and detector) with respect to
the preferred direction of LIV. If this direction is fixed by a vector
$n_{\mu}$ (similar to our previous notations), then, in simplified form, the
Drell--Yan cross section can be written as
\begin{equation}
\frac{d\sigma_{LIV}}{dM^{2}}=\frac{d\sigma_{LI}}{dM^{2}}(1+\delta_{i}%
\frac{E_{i}^{2}}{M^{2}}f_{i}(\Omega,n))
\end{equation}
where $\Omega$ symbolically denotes the orientation of the process in space,
$f_{i}$ are specific functions of $\Omega$ and $n_{\mu}$ and $\delta_{i}$ are
small LIV parameters (not related to any parameters introduced in previous
sections). Here $E_{i}$ is the particle energy and $M$ is so-called invariant mass.

Since the Earth rotates --- and with it the LHC --- the orientation $\Omega$
changes with time, so an anisotropy should appear in the measured cross
sections. Although this result was calculated for a specific type of LIV
modification, the functional form is rather general and applies to a broad
class of LIV operators, including those of the type (\ref{int1}) in leading
approximation. The constraints $\delta_{i}<10^{-5}$ obtained from
cross-section analyses can thus be generalized to any such $\delta_{i}$
provided there is no significant energy dependence.

The prefactor $E_{i}^{2}/M^{2}$, while indicating the expected scaling of LIV
effects with energy, is not a large ratio in this analysis. This is partly
because the final-state particles typically carry significantly less energy
than the full center-of-mass energy, and partly because events with sizeable
$E_{i}^{2}/M^{2}$ are rare and require especially careful treatment. Even
then, only small pseudo-rapidity regions $\left\vert \eta\right\vert <2.4$
were considered. Improving on these points could yield much better sensitivity
to LIV effects, since values like \ $\delta_{i}\sim10^{-5}$ may already be
unrealistically large. This argument is supported by loop calculations in this
and previous work \cite{Loop, Satunin}, although many explicit computations
were performed in simplified QED settings, the qualitative parametric
behaviour --- namely that a loop-generated coefficient is suppressed by a
factor $\dfrac{\alpha}{4\pi}$ relative to a tree-level parameter --- is
generic and continues to hold in the presence of masses. For massive
intermediate bosons this loop factor controls the induced coefficient, while
threshold and finite-mass effects modify only the numerical prefactor and the
scale dependence (logs such as $\ln\frac{M_{LIV}}{M_{B}}$). So, general
outcome holds and LIV effects are transmitted to other sectors, with LIV
parameters suppressed by loop factors (proportional to the relevant
fine-structure constants). Because the weak and QED fine-structure constants
are numerically close, any LIV parameter as large as $\delta_{i}\sim10^{-5}$
would induce unacceptably large corrections in the weak sector and thereby in
all stable particles.

In light of these considerations, the most promising chance for detecting LIV
effects lies in massive intermediate bosons, since they carry the largest
energies in scattering processes. If LIV modifies the dispersion relation of
such a boson --- as in (\ref{photon})--- the consequences for resonance
physics are pronounced:
\begin{equation}
p_{\alpha}^{2}+\delta_{B}(n^{\lambda}p_{\lambda})^{2}=M_{B}^{2} \label{boson}%
\end{equation}
where $M_{B}$ is the boson mass. Although this form looks directly motivated
by (\ref{photon}), it is in fact one of only two renormalizable modifications
compatible with a fixed $n_{\mu}$. The alternative,
\begin{equation}
p_{\alpha}^{2}+m_{B}(n^{\lambda}p_{\lambda})=M_{B}^{2}%
\end{equation}
with small LIV parameter $m_{B}$, produces a much weaker energy scaling and
is therefore harder to detect. At the same time accepting (\ref{photon}),
makes its easy to consider the form (\ref{boson}) for the weak bosons
specifically (with a cartain level of fine-tuning), because whatever origin of
LIV may be photon is still mixed state of $U(1)_{y}$ and $SU(2)$ gauge bosons.

Proceeding with this choice, the resonance behavior is significantly modified,
since both the resonance mass and width are governed by
\begin{equation}
p_{\alpha}^{2}=M_{B,eff}^{2}=M_{B}^{2}-\delta_{B}(n^{\lambda}p_{\lambda})^{2}
\label{Meff}%
\end{equation}

This effective mass picture has immediate experimental consequences. For
example, if the decay rate is modified as \cite{Chkareuli-Kep},
\begin{equation}
\Gamma_{LIV}=\frac{M_{B,eff}^{2}}{M_{B}^{2}}\Gamma_{LI}%
\end{equation}
then the unstable boson propagator becomes
\begin{align}
D_{B}  &  =\dfrac{i}{p_{\alpha}^{2}-M_{B}^{2}}\rightarrow\frac{i}{p_{\alpha
}^{2}-(M_{B,eff}-ip_{0}\Gamma_{LIV}/2M_{B,eff})^{2}}\nonumber\\
&  =\frac{i}{p_{\alpha}^{2}-M_{B,eff}^{2}(1-i\Gamma_{LI}/2M_{B})^{2}}%
\end{align}

Consequently, the scattering cross section involving this intermediate boson
is proportional to:%
\begin{equation}
\sigma_{B}^{_{LIV}}\sim\left\vert D_{B}\right\vert ^{2} \label{LIV distr.}%
\end{equation}

From this, we find for the resonance mass and peak cross section:%
\begin{align}
M_{res}^{2}  &  =M_{B,eff}^{2}(1-\Gamma_{LI}^{2}/4M_{B}^{2})\\
\sigma_{B\max}^{LIV}  &  =\sigma_{B\max}^{LI}\frac{M_{B}^{2}}{M_{B,eff}^{2}%
}\text{ , \ since }\sigma_{B\max}^{LI}\sim\frac{1}{\Gamma_{LI}}%
\end{align}

Applying this framework to the weak Z boson and comparing the resonance mass
shift with the current precision of $M_{Z}$, $\left\vert \Delta M_{Z}%
\right\vert =\left\vert M_{Z}-M_{Zresonance}\right\vert \approx2$ MeV
\cite{Zmass}, using\emph{ }(\emph{\ref{Meff}}) for timelike\emph{ }%
$n^{\lambda}$, we estimate at LHC energies, $E=14$ TeV
\begin{equation}
\left\vert \delta_{B}\right\vert \approx\frac{2M_{Z}\left\vert \Delta
M_{Z}\right\vert }{E^{2}}\approx2\cdot10^{-9} \label{est}%
\end{equation}

If LIV is present, then the data actually follow the LIV-modified distribution
$\sigma_{B}^{_{LIV}}$. Fitting the same data under the LI assumption still
yields a fit, but with shifted parameters. For example, generating data with
$\delta_{B}=-2\cdot10^{-9}$ at energies 14 TeV--- using the Z boson mass and
decay rate --- and fitting it with the LI form results in a shift of $\Delta
M_{Z}\approx-2$ MeV (expected number according to (\ref{est})) in the
resonance peak of the perceived LI cross section, while the decay rate remains
largely unaffected.

The sign and magnitude of this shift depend both on the sign of $\delta_{B}$
and on the energy of the process. To demonstrate the difference between LI and
LIV cross sections, we can plot the relative deviation $\Delta\sigma
/\sigma=(\sigma^{LIV}-\sigma^{LI})/\sigma^{LI}$. This provides a robust
comparison, as it does not rely on the exact form of the cross section but
rather on the ratio of LIV to LI propagators. Thus, we obtain:
\begin{figure}[H]
\centering
\includegraphics[width=1\linewidth]{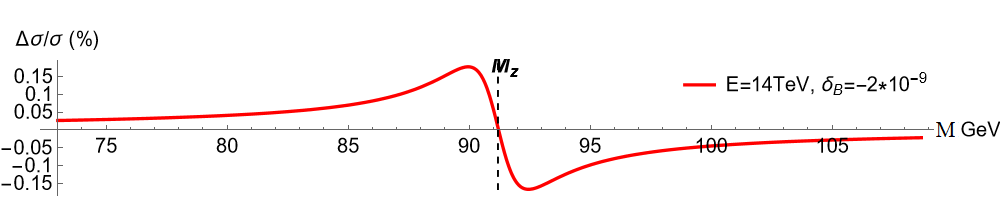} \caption{Relative deviation
$\Delta\sigma/\sigma$ as a function of the invariant mass of the $Z$ resonance
is plotted for $E=14$ TeV and $\delta_{B}=2\times10^{-9}$. The vertical dashed
line marks the $Z$-boson mass $M_{Z}$. LIV effects are nonzero at $M_{Z}$ and
reach a maximum of about $0.17\%$ around 1.2 GeV away from the $M_{Z}$ value.}%
\label{fig:placeholder}%
\end{figure}

This shows that LIV effects are maximal near the resonance region, where the
LIV deformation competes with the mass term, and decrease away from it, where
LIV scales like the dominant contribution and gives only a fractional
correction of order\emph{ }$\delta_{B}$\emph{.}

The resonance line shape also depends on the total center-of-mass energy
through the effective mass of the intermediate boson. At larger energies, the
deformation becomes more pronounced, and the perceived resonance position is
shifted further from the true mass. For processes with lower center-of-mass
energy--close to the boson mass---the resonance aligns with the real mass,
and LIV distortions become negligible.

In proton--proton collisions, the intermediate gauge bosons are produced with
a broad distribution of energies defined by parton distribution functions.
High-energy bosons, which would exhibit stronger LIV effects, are produced
much less frequently, leading to a dilution of the observable LIV signal when
all events are combined. To make the picture more explicit, we rewrite the
kinematics in the standard Drell--Yan variables. Using the invariant
mass\emph{ }$M$\emph{ }and boson rapidity\emph{ }$Y$\emph{, }the momentum can
be parametrized as\emph{ }%
\begin{equation}
p_{\alpha}=M(\cosh Y,\overrightarrow{r}\sinh Y),\text{ \ \ \ }p_{\alpha
}p^{\alpha}=M^{2}%
\end{equation}
where vector\emph{ }$\overrightarrow{r}$\emph{ }defines an axis of collision.
For a timelike\emph{ }$n^{\lambda}$\emph{ }the effective LIV-modified mass
becomes\emph{ }%
\begin{equation}
M_{B,eff}^{2}=M_{B}^{2}-\delta_{B}M^{2}\cosh^{2}Y
\end{equation}
and the position of the resonance shifts to
\begin{equation}
M_{res}^{2}\approx M_{B}^{2}(1-\delta_{B}\cosh^{2}Y)-\Gamma_{LI}^{2}/4
\end{equation}

Thus, the deformation grows approximately like\emph{ }$\cosh^{2}Y$, and
reaches its largest values at high rapidity. At the LHC, rapidities up
to\emph{ }$Y\approx5$\emph{ }are kinematically allowed (defined from the\emph{
}$Y_{\max}=\ln(E/M_{z})$\emph{, }but events with such large Y are rare. This
naturally motivates a search strategy based on rapidity (energy) segregation:
low-rapidity samples determine the physical mass, while high-rapidity subsets
isolate potential LIV-induced deviations. Given the rapidity-dependent nature
of LIV-induced resonance shifts, screening high-rapidity processes near the
resonance region provides a distinct signature that can be disentangled from
Standard Model backgrounds, which typically lack such specific kinematic dependence.

This suggests that accelerator experiments, if analyzed carefully, can be more
sensitive to LIV than previously thought --- potentially on par with
cosmological bounds on neutrinos, $\Delta c/c\sim10^{-8}$ $-10^{-9}$\emph{.}

Of course, this is not a full analysis and a more detailed investigation
involving strict calculations of Drell-Yan processes is needed for definitive
conclusions. Nevertheless, this preliminary estimate highlights a promising
avenue for LIV searches, suggesting that collider studies may offer
constraints comparable to or even surpassing astrophysical observations. This
is even more interesting in the context of recent tension between measurements
of the weak $W$-boson mass by Tevatron and LHC, being outside of each other's
range of accuracy \cite{CDF, LHC}. It is unclear yet whether this tension is
genuine or product of the mistake or bias, but if there is a merit to it, it
is consistent with LIV behavior.

\section{Conclusion}

Lorentz invariance is a foundational principle of modern physics, yet the
possibility of its violation remains an intriguing direction for theoretical
exploration and experimental investigation. In this work, we analyzed how LIV,
when introduced into one sector of the theory, propagates through loop
corrections, affecting particle propagation and dispersion relations. Using
self-energy and vacuum polarization graphs, we demonstrated how LIV influences
kinematics, leading to measurable effects even when initially suppressed.

Experimental constraints from cosmic-ray observations, neutrino astrophysics,
and high-energy collider studies impose limits on these induced LIV
parameters. While direct interaction-based LIV effects would require
unrealistically large values for detection, modifications to dispersion
relations offer a more promising route. In particular, accelerator-based
resonance studies can probe LIV down to $\delta\sim10^{-8}-10^{-9}$. A full
process-specific cross-section calculation is an interesting next step and is
currently being considered; however, the resonance-shift effect described here
is essentially model-independent and therefore provides a robust target for
collider studies.

Our findings emphasize the interconnected nature of LIV effects, showing that
even when introduced in a specific sector, they inevitably spread across
others via quantum loops. This underlines the importance of considering
secondary effects when evaluating LIV constraints. Looking ahead, a deeper
examination of Drell--Yan processes and precision resonance measurements at
future collider experiments could further refine these bounds and provide
valuable insights into the viability of LIV in fundamental physics.

\section{Acknowledgments}

This research is financed by SRNSF \emph{(grant number: STEM-22-2604)}. I
would like to thank Jon Chkareuli and Juansher Jejelava for useful discussions.

\bigskip\appendix

\section*{Appendix A: Gauge invariance breaking in the modified photon
polarization loop}

The polarization diagram modified by the LIV axial mass term is given as
follows:
\begin{equation}
\Pi^{\mu\nu}(p;b)=\int\frac{d^{4}q}{(2\pi)^{4}}\,\mathrm{Tr}[\gamma^{\mu
}\,S(q)\,\gamma^{\nu}\,S(q-p)],
\end{equation}
with $S(k)=1/(\not k  -\not b  \,\gamma^{5})$. We can assume $b^{\nu}$ to be
small and expand in the series with respect to $b^{\nu}$ up to the second
order. We also neglect the mass term for the fermion. These assumptions will
have no bearing on the conclusion. So,
\begin{align}
S(k)  &  \approx S^{(0)}(k)+S^{(1)}(k)+S^{(2)}(k)\label{proexp}\\
S^{(0)}(k)  &  =\frac{1}{\not k  },\quad S^{(1)}(k)=\frac{1}{\not k
}\,(\not b  \,\gamma^{5})\,\frac{1}{\not k  },\quad S^{(2)}(k)=\frac
{1}{\not k  }\,(\not b  \,\gamma^{5})\,\frac{1}{\not k  }\,(\not b
\,\gamma^{5})\,\frac{1}{\not k  }%
\end{align}
Thus%
\[
\Pi^{\mu\nu}(p;b)\approx\Pi_{0}^{\mu\nu}+\Pi_{b}^{\mu\nu}+\Pi_{bb}^{\mu\nu}%
\]
$\ \Pi_{0}^{\mu\nu}$ is standard GI ($\Pi_{0}^{\mu\nu}p_{\mu}=0$)\ result.
$\Pi_{b}^{\mu\nu}$ corresponds to a Chern--Simons--like term, which can appear
GI in form ($\Pi_{b}^{\mu\nu}p_{\mu}=0$). But $\Pi_{bb}^{\mu\nu}$ is not GI
any more even in form. If we calculate
\begin{align}
\Pi_{bb}^{\mu\nu}  &  =\int\frac{d^{4}q}{(2\pi)^{4}}\,\mathrm{Tr}[\gamma^{\mu
}\,S^{(1)}(q)\,\gamma^{\nu}\,S^{(1)}(p-q)\\
&  +\gamma^{\mu}\,S^{(2)}(q)\,\gamma^{\nu}\,S^{(0)}(p-q)+\gamma^{\mu}%
\,S^{(0)}(q)\,\gamma^{\nu}\,S^{(2)}(p-q)]\nonumber
\end{align}
we can check whether $p_{\mu}p_{\nu}\Pi_{bb}^{\mu\nu}$ is zero or not. If it
is not a zero, then GI is broken. Only general form $\Pi_{bb}^{\mu\nu}$ can
assume is the following%
\begin{equation}
p_{\mu}p_{\nu}\Pi_{bb}^{\mu\nu}=A(p_{\mu}^{2})p_{\nu}^{2}b_{\lambda}%
^{2}+B(p_{\nu}^{2})(p^{\mu}b_{\mu})^{2}%
\end{equation}
while calculation of $A(p_{\mu}^{2})$ and $B(p_{\nu}^{2})$ is straightforward,
it is rather lengthy. Therefore, we can simplify our task by putting $p_{\nu}$
on the mass shell ($p_{\nu}^{2}=0$). If GI is conserved, it should remain
conserved on the mass shell as well. Applying the on-shell condition and
symmetry properties of the integrals gives
\begin{equation}
p_{\mu}p_{\nu}\Pi_{bb}^{\mu\nu}=B(p_{\nu}^{2})(p^{\mu}b_{\mu})^{2}=\frac
{16}{3}(p^{\mu}b_{\mu})^{2}\int\frac{d^{4}q}{(2\pi)^{4}}\frac{1}{\left(
q_{\mu}^{2}\right)  ^{2}}%
\end{equation}
which shows explicitly that GI is broken.

\end{document}